\documentstyle[12pt,epsfig]{article}
\author{Juli\'an Candia$^1$ and Ezequiel V. Albano$^{1,2}$\\ 
{\small\it 1 - Instituto de F\'{\i}sica de L\'{\i}quidos y Sistemas 
Biol\'ogicos,}\\{\small\it (CONICET, UNLP), 59 Nro 789, 1900 La Plata, Argentina}\\
{\small\it 2 - Departamento de F\'{\i}sica (UNLP), La Plata, Argentina}}
\title{Nonequilibrium critical behavior of magnetic thin films grown in a temperature gradient}

\begin{document}
\maketitle

\begin{abstract}
We investigate the irreversible growth of $(2+1)-$dimensional magnetic thin films under the 
influence of a transverse temperature gradient, which is maintained by thermal baths across 
a direction perpendicular to the direction of growth. Therefore, different longitudinal layers grow 
at different temperatures between $T_1$ and $T_2$, where $T_1<T_c^{hom}<T_2$ and $T_c^{hom}=0.69(1)$ is 
the critical temperature of films grown in homogeneous thermal baths. 
We find a far-from-equilibrium continuous order-disorder phase transition 
driven by the thermal bath gradient. We characterize this gradient-induced critical behavior 
by means of standard finite-size scaling procedures, which lead to the 
critical temperature $T_c=0.84(2)$ and a new universality class consistent with 
the set of critical exponents $\nu=3/2$, $\gamma=5/2$, and $\beta=1/4$.
In order to gain further insight into the effects of the temperature gradient, 
we also develop a bond model that captures the magnetic film's growth dynamics.  
Our findings show that the interplay of geometry and thermal bath asymmetries leads to growth bond flux 
asymmetries and the onset of transverse ordering effects that explain qualitatively 
the shift observed in the critical temperature. The relevance of these mechanisms is further confirmed 
by a finite-size scaling analysis of the interface width, which shows that the growing sites of the system define a self-affine interface.
\end{abstract}

\section{Introduction}
The importance of thin film technology has been widely recognized in the realms of experimental and applied science, 
from the manufacture of electronics (layers of insulators, 
semiconductors, and conductors from integrated circuits) to optics (reflective 
and anti-reflective coatings, self-cleaning glasses, etc) and packaging (e.g. aluminium-coated PET films).  
Indeed, the increasing role of thin films in basic and applied research relies on the development and refinement of 
nanoscale deposition techniques such as sputtering and molecular beam epitaxy, which allow a single 
layer of atoms to be deposited at a time~\cite{buns94,maha00,ohri02,gloc10}. 

Since the growth temperature is one of the critical parameters in the formation of ordered thin films, 
several experiments have focused on the influence of a temperature gradient during film growth.
In an early experiment by Tanaka {\it et al.}~\cite{tana87}, magnetic Tb-Fe films were grown between two 
substrates with a temperature gradient, reporting the observation of perpendicular 
magnetic anisotropies and other gradient-driven structural features. More recently, 
Schwickert {\it et al.}~\cite{schw00} introduced the ``temperature wedge method'' 
where a calibrated 
temperature gradient of several hundred Kelvin was established across the substrate during co-deposition 
of Fe and Pt on MgO(001) and MgO(110) substrates. These experiments generated the $L1_0$ ordered phase 
of FePt, which is currently the leading candidate material for ultrahigh density heat assisted magnetic 
recording (HAMR) and bit patterned magnetic recording (BPMR) media (\cite{seem07,wang11} and references 
therein). Other experiments by Yongxiong {\it et al.}~\cite{yong05} have investigated the evolution of Fe oxide 
nanostructures on GaAs(100) by using a multi-technique experimental setup that included 
transmission and reflection high energy 
electron diffraction and scanning electron microscopy. In these studies, nanoscale epitaxial 
Fe films were grown, oxidized, and annealed using a gradient temperature method, which led to 
nanostripes with uniaxial magnetic anisotropy. 
As a result of the experimental advances on this field, many technological applications have been envisioned as well. 
For instance, magneto-optical recording studies of signal reproduction~\cite{kane00} have suggested that 
recording media having multiple magnetic layers in a transverse 
temperature gradient may suppress magnetic noise from tracks adjacent to the target track during information 
storage and reproduction~\cite{kawaguchi}.  

From a theoretical perspective, gradients have been studied extensively in the context of diffusion 
processes and later extended to thermal conductivity and heat conduction problems. 
The so-called gradient percolation method was originally introduced to 
study percolation transition models where the density is the control parameter \cite{sapo85} and later applied 
to a variety of problems, such as fractal diffusion fronts~\cite{mems00,hade02,chap04}, 
overlapping disks in a concentration gradient~\cite{ross89}, 
bond percolation for the Kagom\'e lattice~\cite{ziff97}, invasion percolation under gravity~\cite{gouy05}, 
porous media~\cite{ross86}, as well as in the study of vegetation distribution~\cite{gast09}. 
Very recently, the gradient method has been extended as a powerful tool to study first- and second-order 
irreversible phase transitions in far-from-equilibrium systems such as the Ziff-Gulari-Barshad model 
and forest-fire cellular automata~\cite{losc09,guis11}.   
In magnetic systems, damage spreading processes in a temperature gradient~\cite{bois91} 
and studies of several one-dimensional models~\cite{sait96,solo01,savi05,leco05} 
have been followed by the 
investigation of the kinetic Ising model in two dimensions under a variety of 
dynamics~\cite{sait99,neek06,casa07,agli07,mugl11}.

Within the broad context of these recent experimental and theoretical investigations, the aim of this 
paper is to study the irreversible growth of magnetic thin films 
in a temperature gradient and to provide a full characterization of the gradient-induced critical phase transition.  
The magnetic thin film growth process under far-from-equilibrium conditions 
is investigated by using the so-called magnetic Eden model (MEM)~\cite{ausl93,cand01,cand11}, an extension 
of the classical Eden model~\cite{eden58} in which particles 
have a two-state spin as an additional degree of freedom. 
Earlier studies have shown that, 
growing in $(d+1)$-dimensional strip geometries in homogeneous thermal baths, MEM films are 
noncritical for $d=1$~\cite{cand01}. 
In contrast, for $d=2$ they undergo an order-disorder phase transition that takes place at 
$T_c^{hom}=0.69(1)$ in the thermodynamic limit. The critical exponents are $\nu^{hom}=1.04(16)$, 
$\gamma^{hom}=2.10(36)$, and $\beta^{hom}=0.16(5)$, which intriguingly agree within error 
bars with the exact exponents for the kinetic Ising model~\cite{cand01}. 
Since the MEM growth process 
is irreversible and newly deposited particles are not allowed to flip and thermalize once they 
are added to the growing cluster, the observed correspondence between the MEM and the equilibrium Ising 
model remains puzzling.  

In this work, we focus on the critical case (i.e. $d=2$) and show that, when applying a transverse 
temperature gradient maintained by thermal baths between temperatures 
$T_1$ and $T_2$, where $T_1<T_c^{hom}<T_2$, the system undergoes a continuous phase transition 
at a higher critical temperature ($T_c>T_c^{hom}$) and with different critical exponents. 
We also develop a growth bond model and show the existence of bond flux asymmetries caused by the 
interplay of geometry and thermal bath asymmetries, which shed some light on the growth dynamics 
and explain qualitatively the shift observed in the critical temperature. The growth bond model 
analysis is further supported by the fact that the growing interface is self-affine, thus ensuring 
that the growing sites are correlated at all size scales.

The rest of the paper is laid out as follows. In Section 2, we introduce the model and 
describe the Monte Carlo algorithm used to simulate MEM thin films in a temperature gradient. 
In Section 3, we present our results and a discussion. Finally, Section 4 consists of concluding remarks.

\section{The Model and the Monte Carlo Simulation Method}
The MEM in $(2 + 1)-$dimensions is studied in the square lattice 
by using a rectangular geometry $L_x \times L_y\times L_z$, where $L_z\gg L_x=L_y\equiv L$ 
is the growth direction. The location of each spin on the lattice is specified through its 
coordinates $(x,y,z)$ ($1 \leq x,y \leq L$, $1 \leq z \leq L_z$). 
The starting seed for the growing cluster is a plane of $L \times L$ 
parallel-oriented spins placed at $z=1$ and cluster growth takes place 
along the positive longitudinal direction (i.e., $z\geq  2$).
Across one of the transverse directions (the $y-$axis), a temperature gradient is applied by 
thermal baths at fixed temperatures linearly varying between $T_1$ and $T_2$. 
Therefore, in our setup each layer at fixed $y$ is subjected to a constant 
temperature $T(y)=T_1+(T_2-T_1)(y-1)/(L-1)$ maintained by a thermal bath. 
We adopt open boundary conditions along the $y-$direction, while continuous boundary 
conditions are considered along the $x-$direction. 

Clusters are grown by selectively adding two-state spins ($S_{xyz}= \pm 1$) to perimeter sites, 
which are defined as the nearest-neighbor (NN) empty sites of the already occupied ones. 
Let us recall that the substrate is a $3D$ cubic lattice and therefore each lattice 
site in the bulk has 6 NN sites. 
Considering a ferromagnetic interaction of strength $J>0$ between NN spins, 
the energy $E$ of a given configuration of spins is given by 
\begin{eqnarray}
E=-\frac{J}{2}\sum_{\langle xyz,x'y'z'\rangle}
S_{xyz}S_{x'y'z'}\  ,
\label{energy}
\end{eqnarray}
\noindent where the summation $\langle xyz,x'y'z'\rangle$ 
is taken over occupied NN sites.  
The Boltzmann constant is set equal to unity throughout, 
and both temperature and energy are measured 
in units of $J$. The probability for a perimeter site at $(x,y,z)$ 
to be occupied by a spin is proportional to the Boltzmann factor
$\exp(- \Delta E/{T})$, where $\Delta E$
is the change of energy involved in the addition of
the spin and $T$ is the temperature at the perimeter site. 

At each step, all perimeter sites have to be considered and the probabilities 
of adding a new (either up or down) spin to each site must be evaluated. 
Using the Monte Carlo simulation method, after all probabilities are 
computed and normalized, the growth site and the orientation of the 
new spin are both simultaneously determined by means of a pseudo-random number.
Notice that the MEM's growth rules require updating the probabilities at 
each time step and lead to very slow algorithms compared with analogous equilibrium spin models.
Since the observables of interest (e.g. the mean transverse magnetization along the $x-$direction and its 
higher moments) require the growth of samples with a large number of transverse planes of size $L\times L$, 
clusters having up to $10^9$ spins have typically been grown for lattice sizes in the range $12\leq L\leq 96$.
Also, let us point out again that, although Eq.~(\ref{energy}) resembles the Ising Hamiltonian, 
the MEM is a nonequilibrium model in which new spins are continuously added, 
while older spins remain frozen and are not allowed to flip, detach, nor diffuse.
 
As in the case of the classical Eden model, the magnetic Eden model 
leads to a compact bulk and a self-affine growth interface~\cite{ausl93} (see Sect. 3.4 for a detailed 
finite-size scaling analysis of the interface width). 
The growth front may temporarily create voids within the bulk, usually not far from the rough growth interface. 
However, since the boundaries of these voids are also perimeter sites, they ultimately 
become filled at some point during the growth process. 
Hence, far behind the active growth interface, the system is compact 
and frozen, and the different quantities of interest can thus be measured on defect-free transverse planes.  

Notice that the growth of magnetic Eden aggregates in $(2+1)$-strip geometries is 
characterized by an initial transient 
length $\ell_T\sim L^2$ (measured along the growth direction, i.e. the $z-$axis) 
followed by a nonequilibrium stationary state that is independent of the initial 
configuration~\cite{cand01}. We considered starting seeds formed by $L\times L$ up spins (i.e. $S_{xy1}=1$) 
but any choice for the seed leads to the same stationary states for $z\gg\ell_T$.  
By disregarding the transient region, all results reported in this paper are 
obtained under stationary conditions.
Also notice that, since the films are effectively semi-infinite and the substrate length along the 
growth direction plays no role, the only characteristic length is the transverse linear 
size $L$.    

\section{Results and Discussion}

\subsection{Gradient-Driven Continuous Pseudo-Phase Transitions in Finite-Size Films}

\begin{figure}[t!]
\centerline{{\epsfxsize=5.2in \epsfysize=2.1in \epsfbox{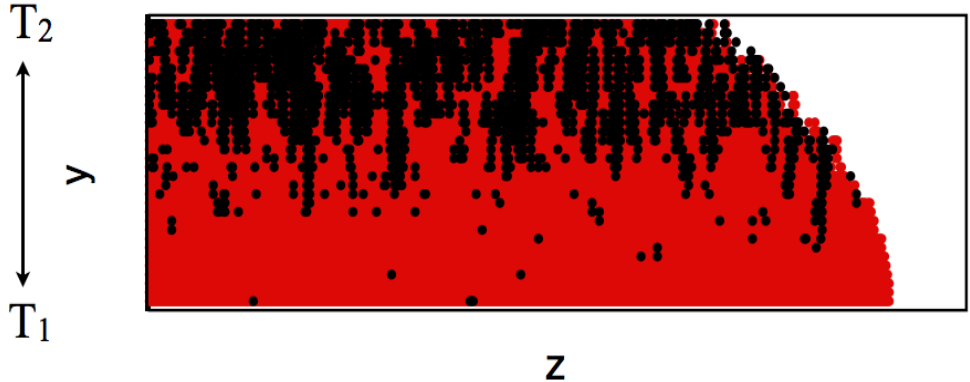}}}
\caption{Snapshot showing a longitudinal slice for a fixed value of the transverse coordinate $x$. 
A temperature gradient between $T_1=0.5$ (bottom) and $T_2=1.5$ (top) 
is maintained along the transverse axis $y$. 
The system grows along the longitudinal $z>0$ direction in a semi-infinite $(2+1)$-dimensional 
strip substrate. Red (black) sites represent up (down) spins, while empty sites are shown in white.}
\label{fig1}
\end{figure}

Let us begin by considering a fixed gradient between temperatures $T_1=0.5$ and $T_2=1.5$. The 
effect of changing this gradient will be discussed later.
Figure 1 shows the snapshot of a longitudinal slice for a fixed value of the transverse coordinate $x$.
The temperature gradient is applied along the transverse $y-$direction, while the system grows from 
left to right along the longitudinal $z>0$ direction. Notice that the bulk grows compact, 
because although voids and holes in the bulk may eventually occur, they ultimately become filled at some 
point during the growth process. The bottom layers grow in contact with thermal baths 
at cold temperatures, which favor the formation of well-ordered spin domains. In contrast, the top 
layers grow in contact with hot thermal baths that promote bulk disorder. As will be shown below, the 
interplay of model growth dynamics, geometry, and thermal bath asymmetries lead to the onset of 
gradient-driven order-disorder critical phase transitions that can be quantitatively characterized. 

\begin{figure}[t!]
\centerline{{\epsfysize=3.3in \epsfbox{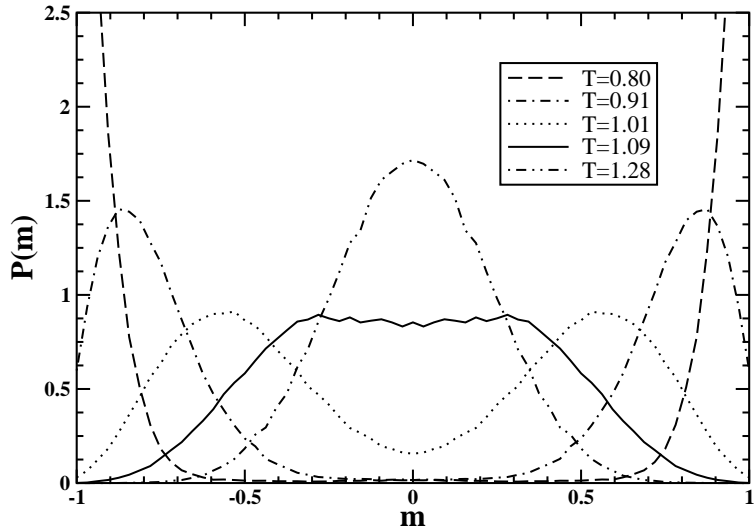}}}
\caption{Symmetrized magnetization probability distributions for a system of linear size $L=64$ 
and different layer temperatures between $T_1=0.5$ and $T_2=1.5$. The sharp peaks near $m\simeq\pm 1$ for $T=0.80$ have been truncated.}
\label{fig2}
\end{figure}

In order to take into account the asymmetries introduced by the temperature gradient, we can 
quantify the degree of order in the system by considering the magnetization of transverse columns 
at constant temperature (i.e. along the $x$-axis): 
\begin{eqnarray}
m(y)=\frac{1}{L}\sum_xS_{xyz}\ . 
\label{magnetization}
\end{eqnarray}
  
Figure 2 shows the probability distributions of $m$ for a system of linear size $L=64$ growing in 
a temperature gradient between $T_1=0.5$ and $T_2=1.5$. Notice that different plots correspond 
to different layers and, therefore, to different temperatures within the gradient's range, as indicated. 
As expected for a continuous order-disorder phase transition, 
the low-temperature distributions are bimodal and peaked at the spontaneous magnetization 
$m=\pm m_{sp}$ $(0 < m_{sp} < 1)$. As the temperature is 
increased, the peaks approach each other and merge smoothly, ultimately leading to a Gaussian distribution 
peaked at $m=0$ for high temperatures, which is characteristic of the disordered phase.
Indeed, the smooth shift of the distribution maxima across $T\simeq T_c(L)$, from the 
low-temperature nonzero spontaneous magnetization $m=\pm m_{sp}$ to the high-temperature 
Gaussian centered at $m=0$, is the signature of true 
thermally-driven continuous phase transitions~\cite{bind02}. 

\begin{figure}[t!]
\centerline{{\epsfysize=3.3in \epsfbox{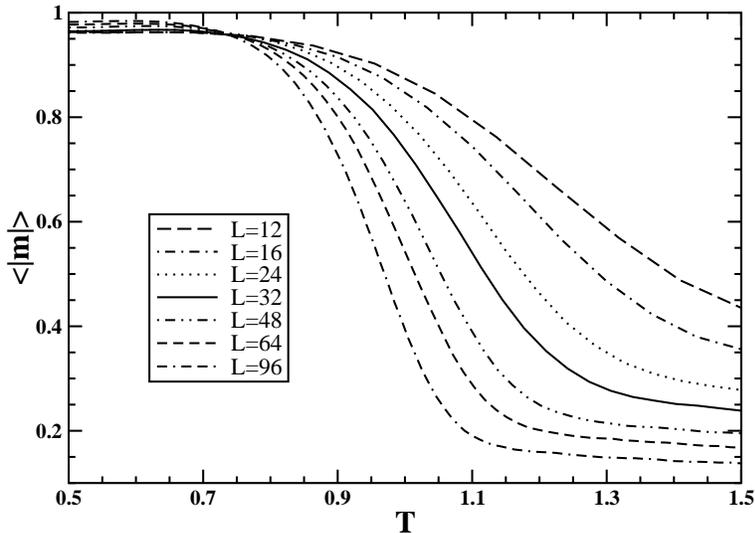}}}
\caption{Mean absolute magnetization as a function of the layer temperature 
for different system sizes, as indicated.}
\label{fig3}
\end{figure}

Notice that the distributions in Figure 2 are symmetrical, since there exists a finite probability for 
fluctuations to grow and switch the magnetization from $m\simeq +m_{sp}$ to $m\simeq -m_{sp}$ and viceversa. 
Since Monte Carlo simulations are restricted to finite samples, the standard procedure to avoid 
these shortcomings due to finite-size effects is to average the absolute value of the order 
parameter~\cite{bind00}. In this context, the appropriate order parameter
is the mean absolute magnetization of transverse columns at constant temperature, i.e. 
\begin{eqnarray}
\langle|m|\rangle(y)=\langle\frac{1}{L}|\sum_xS_{xyz}|\rangle_z\ , 
\label{ordparam}
\end{eqnarray}
\noindent where $\langle ...\rangle_z$ denotes averages along the growth direction $z>0$ within the 
stationary region.
Figure 3 shows plots of the mean absolute magnetization as a function of the layer temperature for different 
system sizes in the range $12\leq L\leq 96$.  
For any given system size, at low temperatures the system grows ordered and the magnetization
is close to unity, while at higher temperatures the disorder sets on and the magnetization  
becomes significantly smaller. 
However, fluctuations due to the finite system size prevent the magnetization from 
becoming strictly zero above the critical temperature, so the transition between the low-temperature 
ordered phase and the high-temperature disordered phase becomes smoothed out and rounded. 
As expected, larger systems are less affected by finite-size effects and display sharper transitions. 

Strictly speaking, however, these results just show evidence of {\it pseudo}-phase transitions, 
which might be precursors of true phase transitions taking place in the thermodynamic limit. 
In the following, we will characterize in more detail this pseudo-critical phenomenon by 
measuring other observables on finite-size systems. In the next Subsection, we will use standard finite-size 
scaling procedures to establish the existence of a non-trivial critical temperature in the $L\to\infty$ limit, 
as well as to calculate critical exponents that describe the 
behavior of the infinite system at criticality. 

\begin{figure}[t!]
\centerline{{\epsfysize=3.3in \epsfbox{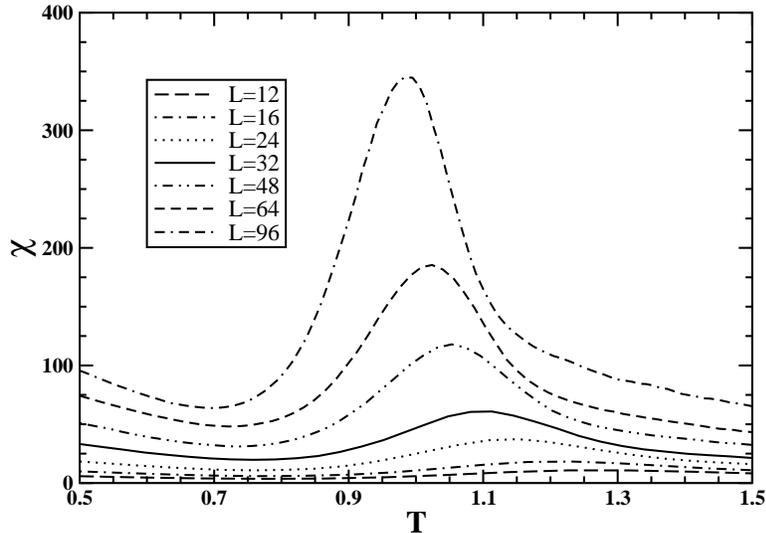}}}
\caption{Susceptibility as a function of the layer temperature for  
different system sizes in the range $12\leq L\leq 96$. 
As expected for a critical system, the peaks become sharper 
and higher as $L$ is increased.}
\label{fig4}
\end{figure}

Let us now consider the magnetic susceptibility $\chi$, given by
\begin{equation}
\chi = \frac{L^2}{T}\left(\langle m^2\rangle-\langle|m|\rangle^2\right).
\label{chimi}
\end{equation}
For equilibrium systems, the susceptibility is related to order parameter fluctuations 
by the fluctuation-dissipation theorem. Although the validity of 
a fluctuation-dissipation relation in the case of a nonequilibrium system
is less evident, we will assume Eq.~(\ref{chimi}) to hold also for the MEM. Indeed, as 
shown in earlier studies of nonequilibrium spin models~\cite{side98,korn01}, 
this definition of $\chi$ proves very useful for exploring critical phenomena under far-from-equilibrium conditions. 
In Section 3.2, we will characterize the critical behavior in the thermodynamic limit through critical exponents 
and finite-size scaling relations by applying the equilibrium theory to our far-from-equilibrium model. 

Figure 4 shows plots of $\chi$ vs $T$ for different system sizes, as indicated. 
As with the thermal dependence of the order parameter shown in Figure 3, the peaks of the susceptibility 
become rounded and shifted, indicating the existence of pseudo-phase transitions in finite-size MEM thin films. 
Increasing the system size, the peaks become sharper and higher, as expected for a critical system. 

\begin{figure}[t!]
\centerline{{\epsfysize=3.3in \epsfbox{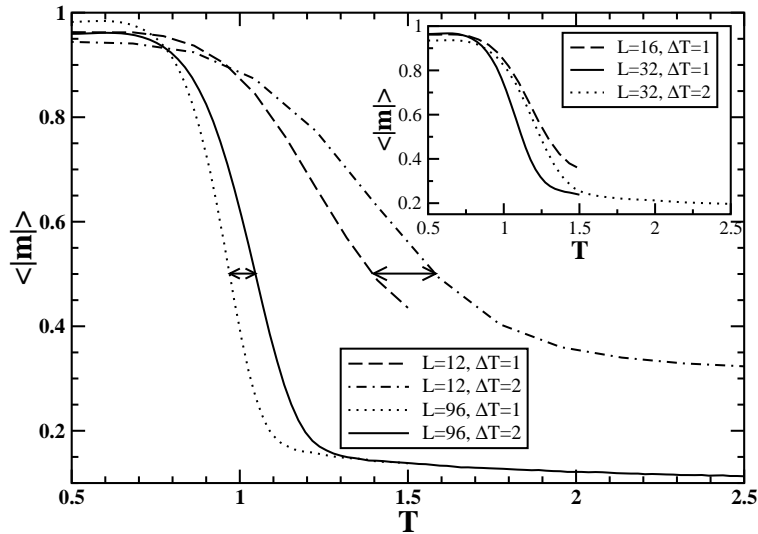}}}
\caption{Comparison of magnetization profiles $\langle |m|\rangle$ vs $T$ for different gradient 
spans ($\Delta T=1,2$) and system sizes ($L=12,96$). The arrows indicate the corresponding 
shifts in the finite-size critical temperature $T_c(L)$.
Inset: Magnetization profiles for systems with the same local gradient $\delta=1/16$, namely 
$L=16$, $\Delta T=1$ (dashed line) and $L=32$, $\Delta T=2$ (dotted line). For comparison, 
the mixed case $L=32$, $\Delta T=1$ (solid line) is also shown.
}
\label{comparison}
\end{figure}

Since the results presented thus far considered a fixed gradient between the 
temperatures $T_1=0.5$ and $T_2=1.5$, let us now investigate the effects of changing the 
gradient span $\Delta T\equiv T_2-T_1$. With this aim, we 
keep the same temperature for the thermal bath at the cold end ($T_1=0.5$) and consider a 
substantially higher temperature for the thermal bath at the hot end ($T_2=2.5$). 

Figure~\ref{comparison} compares the mean absolute magnetization for these two different gradient ranges 
(i.e. $\Delta T=1,2$) for systems of size $L=12$ and $L=96$. 
We observe that increasing the gradient span shifts the magnetization profiles towards 
higher temperatures. That is, the temperature of a given layer does not uniquely determine its degree 
of order, since the mean magnetization of the layer also depends on the overall gradient span 
under which the film grows. Similar shifts towards higher temperatures are 
also seen in higher-order moments of the order parameter probability distributions, such as 
the susceptibility and Binder's fourth-order cumulant (not shown here for the sake of space). 
Alternatively, we can compare systems of different sizes and gradient spans such that the local 
gradients $\delta\equiv\Delta T/L$ are the same. The inset to Figure~\ref{comparison} 
shows a comparison between a system of size $L=16$ and gradient span $\Delta T=1$ (dashed lines) and 
another system of size $L=32$ and gradient span $\Delta T=2$ (dotted lines), 
both of which have the same local gradient $\delta=1/16$. 
The solid line corresponds to the mixed case $L=32$ and $\Delta T=1$ (i.e. $\delta=1/32$).
We observe that the systems with the same local gradient have similar magnetization in layers at intermediate temperatures 
(i.e. approximately in the range $0.8\leq T\leq 1.3$). However,  
the magnetization profiles for equal-$\delta$ systems differ noticeably in layers closer to the borders of the sample.      

The arrows in Figure~\ref{comparison} show that the shifts for smaller systems   
are significantly larger than the corresponding shifts for larger systems. 
Defining the finite-size pseudo-critical temperature $T_c(L)$ as the temperature corresponding to 
$\langle |m|\rangle=0.5$, the shift for $L=12$ is $\Delta T_c=0.18$, while the corresponding shift 
for $L=96$ is $\Delta T_c=0.07$. 
This observation suggests that differences arising from changing the gradient span might 
just reflect finite-size effects that vanish in the $L\to\infty$ limit. In the next 
Subsection, we study the critical behavior of MEM films and confirm 
that, in fact, these differences are merely finite-size effects that become irrelevant in the 
thermodynamic limit. 

\subsection{Characterization of Gradient-Driven Critical Behavior in the Thermodynamic Limit}

So far, we have found evidence for the existence of a gradient-driven order-disorder phase 
transition from the analysis of order parameter probability distributions, the order parameter mean 
absolute value (magnetization) and its fluctuations (susceptibility) in the growth of finite-size 
magnetic films. 
In this Subsection, we apply standard finite-size scaling techniques to show the existence 
of this phase transition in the thermodynamic limit and to determine critical exponents that 
characterize the system's critical behavior and universality class. 

\begin{figure}[t!]
\centerline{{\epsfysize=3.3in \epsfbox{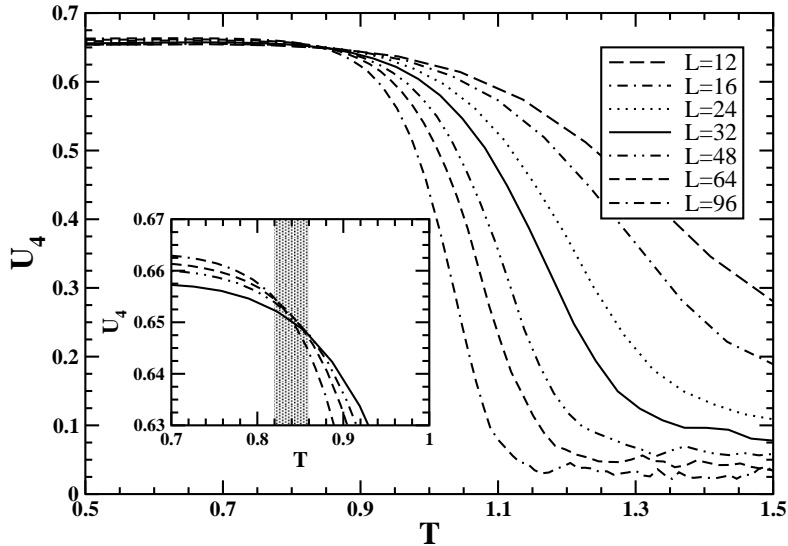}}}
\caption{Binder cumulant as a function of the layer temperature for  
different system sizes, as indicated.
The inset shows the cumulant intersections for the larger systems ($32\leq L\leq 96$), 
which determine $T_c=0.84(2)$.}
\label{cumulant}
\end{figure}

The Binder cumulant, defined by
\begin{eqnarray}
U_4=1-\frac{\langle m^4\rangle}{3\langle m^2\rangle^2}\  ,
\label{binder}
\end{eqnarray}
\noindent is a fourth-order cumulant dependent on the variance and the kurtosis of the order parameter probability 
distribution. 
One important property of the Binder cumulant is that, for large system sizes, the low-temperature, 
ordered region tends to 
the value $2/3$, while the high-temperature, disordered region tends to $0$. Thus, in the thermodynamic 
limit, the function becomes discontinuous exactly at the critical temperature.
Moreover, 
since for second-order phase transitions, the scaling prefactor of the cumulant is independent of the 
sample size, plots of $U_4$ versus the control parameter 
lead to a common (size-independent) intersection point that corresponds to the location of the critical 
value of the order parameter in the thermodynamic limit~\cite{bind81}.  

Figure~\ref{cumulant} shows the Binder cumulant as a function of the layer temperature for different system sizes in the range $12\leq L\leq 96$. 
The Inset to Figure~\ref{cumulant} shows a detailed view of the same data for the largest lattices ($32\leq L\leq 96$), where 
the intersection region is indicated by a grey vertical strip. Based on this observation, we determine 
the critical temperature in the $L\to\infty$ limit as $T_c=0.84(2)$. 
Interestingly, this value is significantly higher than the corresponding critical temperature for 
the MEM growing in an homogeneous thermal bath (i.e. in the absence of a temperature gradient), namely 
$T_c^{hom}=0.69(1)$ \cite{cand01}. In the next Subsection, 
we will explore the growth dynamics and explain qualitatively this 
shift in the critical temperature as due to ordering effects caused by a net transverse growth bond flux 
induced by thermal asymmetries. 

Notice also that, by fixing the temperature range (i.e. the temperatures $T_1$ and $T_2$) and increasing $L$, we are effectively considering different gradients $\delta=(T_2-T_1)/L$ that become smaller as $L$ is increased. Figure~\ref{cumulant}, which shows a fixed point in the Binder cumulants as the gradients are changed, provides therefore quantitative evidence for the existence of a gradient-independent phase transition taking place at the temperature $T_c$. 

According to the finite-size scaling theory, developed for the treatment of finite-size effects 
at criticality under equilibrium conditions~\cite{barb83, priv90},
the difference between the true critical temperature, $T_c$, and the effective 
pseudo-critical one, $T_c(L)$, is given by 
\begin{equation}
|T_c-T_c(L)|\propto L^{-1/\nu}, 
\label{scaling1}
\end{equation}
where $\nu$ is the exponent that characterizes the divergence of the correlation length 
at criticality. As mentioned above, we define the finite-size pseudo-critical temperature 
$T_c(L)$ as the temperature corresponding to $\langle |m|\rangle=0.5$.

\begin{figure}[t!]
\centerline{{\epsfysize=3.3in \epsfbox{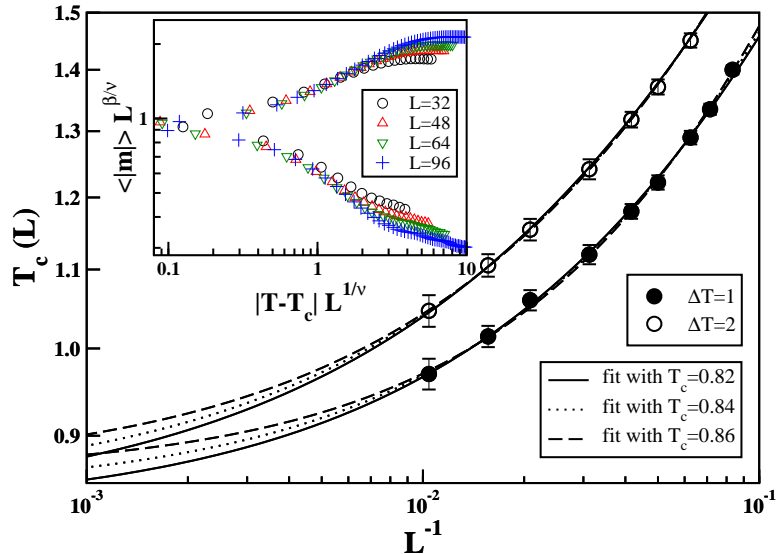}}}
\caption{Log-log plots of the finite-size pseudo-critical temperatures $T_c(L)$ as a function of the 
inverse of the system linear size, $L^{-1}$, for $\Delta T=1$ (filled circles) and $\Delta T=2$ (open circles).
By means of standard finite-size scaling analysis, we obtain the critical exponent $\nu=1.53(6)$ (see more details in the text). Inset: plot of 
$\langle |m|\rangle\times L^{\beta/\nu}$ vs 
$|T-T_c|\times L^{1/\nu}$ (with $\beta=0.26$) 
showing a data collapse for $\Delta T=1$ and different system sizes in the range $32\leq L\leq 96$.}
\label{fig7}
\end{figure}

Let us point out that, given the lack of a comprehensive theory of non-equilibrium phase transitions, 
concepts and definitions developed in the 
context of equilibrium phenomena are customarily borrowed and applied to far-from-equilibrium phenomena as well. For a review of 
standard methods, see e.g.~\cite{marr99,henk09,henk10}. 
Indeed, although this approach is ad-hoc 
and lacks the theoretical foundations of equilibrium systems, it has been used extensively in the literature and has become 
a powerful means of advancing our knowledge within the realm of non-equilibrium phenomena. For instance, numerical methods such as 
Monte Carlo simulations or series expansions are restricted to finite systems and it is therefore 
important to understand how far finite-size 
effects influence the properties of the system. As known from equilibrium statistical mechanics, finite-size effects are particularly 
strong close to the critical point, where the spatial correlation length becomes comparable with the linear dimensions of the system. 
By introducing the system size as an additional parameter, finite-size scaling laws are used to characterize the steady state of 
finite far-from-equilibrium systems through appropriate scaling exponents, such as, for instance, the exponent $\nu$ in Eq.(\ref{scaling1}) above. 
Moreover, this procedure allows us to define universality classes of non-equilibrium systems, as reviewed e.g. in Refs.~\cite{hinr00,odor04,alba11}.

Figure 7 shows log-log plots of the finite-size pseudo-critical temperatures $T_c(L)$ as a function of the 
inverse of the system linear size, $L^{-1}$, for different gradients and system sizes, as indicated. 
By rewriting the finite-size scaling relation as
\begin{equation}
T_c(L)=T_c+A\times L^{-1/\nu}, 
\label{scalingfit}
\end{equation}
we performed different least-squares fits to the data using the mean, upper-bound and lower-bound 
values for the critical temperature in the thermodynamic limit. The nonlinear least-squares fitting procedure was implemented using the Levenberg-Marquardt minimization method~\cite{pres92} and the 
results from each independent fit are reported in Table~\ref{tablefits}. The errors in the table 
are determined by the fitting algorithm and take into account the statistical errors for each 
datapoint. Figure 7 shows that 
the finite-size scaling relation fits the data very well within error bars for the range of values for 
the critical temperature that was derived from the intersection of Binder's cumulants. 
From these fits, we obtain the critical exponent $\nu=1.53(6)$, 
where the error bars reported reflect the errors derived from the evaluation of $T_c$ as well as the statistical errors. 
Notice that the data for different gradients tends to converge in the $L\to\infty$ limit, therefore confirming that 
differences arising from changing the gradient are finite-size effects.  
On the other hand, finite-size
scaling theory predicts that plots of $\langle |m|\rangle L^{\beta/\nu}$ vs 
$|T-T_c| L^{1/\nu}$ for different lattice sizes should collapse near the critical region. 
The inset to Figure 7 shows the data collapse obtained by 
using $\beta=0.26$ (that is determined from the hyperscaling relation, as explained below)
with two separate branches corresponding to the low- and high-temperature regions.  

\begin{table}[t!]
\centering
\begin{tabular}{||c|c|c|c||} \hline \hline
\multicolumn{1}{||c|}{$\Delta T$}&\multicolumn{1}{c|}{$T_c$}&\multicolumn{1}{c|}{$A$}&\multicolumn{1}{c||}{$\nu$}\\ \hline \hline
 1 & 0.82 & 2.86(3) & 1.54(2) \\ \hline
 1 & 0.84 & 2.90(3) & 1.49(2)  \\ \hline
 1 & 0.86 & 2.78(3) & 1.48(1)  \\ \hline
 2 & 0.82 & 3.70(3) & 1.58(1) \\ \hline
 2 & 0.84 & 3.65(3) & 1.56(2)  \\ \hline
 2 & 0.86 & 3.66(4) & 1.53(2)  \\ \hline \hline
\end{tabular}
\
\caption{Results from fitting the data to the finite-size scaling relation, Eq.(\ref{scalingfit}).}
\label{tablefits}
\end{table}

\begin{figure}[t!]
\centerline{{\epsfysize=3.3in \epsfbox{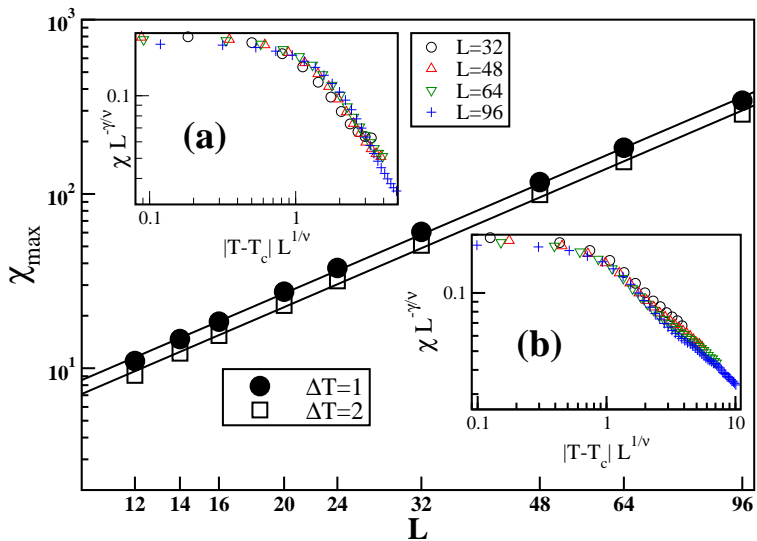}}}
\caption{Log-log plots of the susceptibility maxima as a function of the system linear size
for $\Delta T=1$ (filled circles) and $\Delta T=2$ (open squares), where 
statistical errors for each datapoint are smaller than the symbol size. 
The solid lines are finite-size scaling fits that yield $\gamma/\nu=1.66(3)$. The insets display  
plots of $\chi L^{-\gamma/\nu}$ vs $|T-T_c| L^{1/\nu}$ (for the $\Delta T=1$ case) showing separately 
the data collapse for (a) the low-temperature branch and (b) the high-temperature branch.
}
\label{fig8}
\end{figure}

An additional characterization of the critical behavior of this system can be obtained 
by calculating the critical exponent $\gamma$, which describes the divergence of the susceptibility 
at the critical point. Using again the finite-size scaling theory~\cite{barb83,priv90}, 
the exponent ratio $\gamma/\nu$ is related to the peak of the susceptibility measured in finite samples of size $L$ by
\begin{equation}
\chi_{max}\propto L^{\gamma/\nu}\ .
\label{scaling2}
\end{equation}

The symbols in Figure 8 correspond to the maxima of $\chi$ plotted against the system linear size for different 
gradients, as indicated, while the solid lines are fits to the data using the scaling relation from Eq.(\ref{scaling2}). Statistical errors for each datapoint are smaller than the symbol size in the Figure. 
The fitting procedure (which, as mentioned above, was implemented as a nonlinear least-squares algorithm using the Levenberg-Marquardt minimization method~\cite{pres92}) yields $\gamma/\nu=1.66(3)$, where the error bars reflect the statistical errors from the fits. 
Using this ratio and the value already obtained for $\nu$, we determine $\gamma=2.54(11)$. 
The insets to Figure 8 display plots of 
$\chi L^{-\gamma/\nu}$ vs $|T-T_c| L^{1/\nu}$ 
for $\Delta T=1$ and different lattice sizes in the range $32\leq L\leq 96$. Using the critical temperature as determined by the 
susceptibility peaks, the data collapse is shown separately for (a) the low-temperature branch and (b) the high-temperature branch. In the former case, data from low-temperature layers near $T_1=0.5$ depart from the collapse 
and have been removed. However, the collapse near the critical region is remarkable and agrees very well with the expectations from the finite-size scaling theory.  
By replacing the exponents $\nu$ and $\gamma$ in the hyperscaling relation $d\nu-2\beta-\gamma=0$ with $d=2$, we determine the exponent $\beta=0.26(8)$, where the error is determined from standard 
error propagation applied to the hyperscaling relation. 
Recall that we anticipated this value of $\beta$ when we considered the 
data collapse of the scaled magnetization (see the inset to Figure 7 above). The excellent data collapse 
near the critical region confirms the consistency and robustness of the obtained results. 

As a summary, Binder's cumulant method and finite-size scaling analysis allowed us to characterize quantitatively 
the critical behavior of nonequilibrium magnetic 
films growing in a temperature gradient. We found that differences arising from changing 
the gradient are due to finite-size effects that vanish in the thermodynamic limit. The system's critical temperature 
is $T_c=0.84(2)$, significantly higher than the critical temperature for films grown in an homogeneous 
thermal bath, $T_c^{hom}=0.69(1)$~\cite{cand01}. The critical exponents are $\nu=1.53(6)$,  
$\gamma=2.54(11)$, and $\beta=0.26(8)$. 
Based on our findings, we conjecture that magnetic Eden films growing in a temperature gradient belong to a new universality class 
characterized by critical exponents $\nu=3/2, \gamma=5/2,$ and $\beta=1/4$.   
In contrast, the critical exponents for magnetic Eden films grown in an homogeneous bath agree within error 
bars with the exact exponents for the Ising model in $d=2$~\cite{cand01}, namely $\nu=1, \gamma=7/4$, and $\beta=1/8$.  

\subsection{Growth Bond Model and Bond Flux Asymmetries}

In this Subsection, we explore the growth dynamics by means of a simple
bond representation. Let us recall that the MEM's growth process adds new spins, 
which are deposited one by one to the growing cluster. Although voids and holes may form within 
the bulk, ultimately all sites become filled. Hence, to each pair of neighboring sites, 
we can assign a directed bond that points from the earlier occupied site to the later occupied one. 
The components of the bond flux field $\vec{\phi}$ at a site $(x,y,z)$ are defined as:
$$
\phi_x(x,y,z)=b[(x,y,z),(x+1,y,z)]\ ,
$$
\begin{equation}
\phi_y(x,y,z)=b[(x,y,z),(x,y+1,z)]\ ,
\label{bondflux}
\end{equation}
$$
\phi_z(x,y,z)=b[(x,y,z),(x,y,z+1)]\ , 
$$
where $b[s_1,s_2]=1$ if the bond points from $s_1$ to $s_2$, and $b[s_1,s_2]=-1$ if the bond points from $s_2$ to $s_1$.  

\begin{figure}[t!]
\centerline{{\epsfysize=3.3in \epsfbox{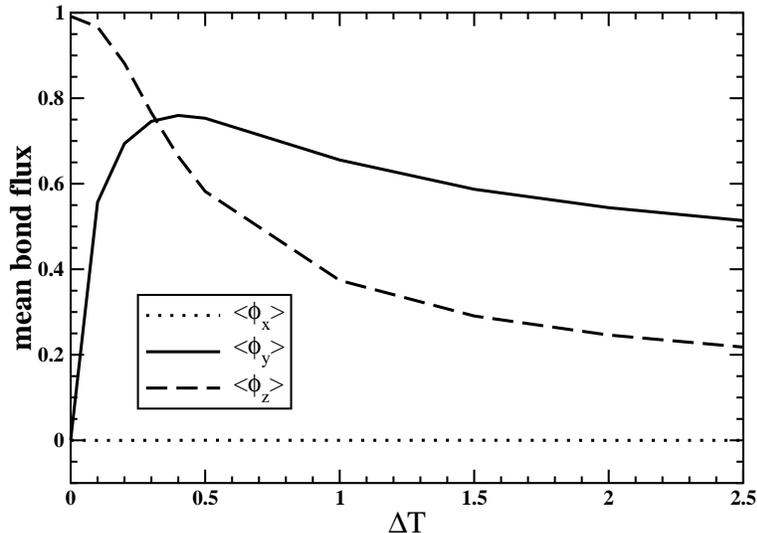}}}
\caption{Mean growth bond flux components as a function of the gradient span 
$\Delta T$ for $L=32$ and $T_1=0.5$. Asymmetries due to the temperature 
gradient and the substrate geometry are responsible for net bond fluxes along 
the $y$ and $z$ directions. Along the transverse direction $x$ the system is fully symmetric, 
so no net bond fluxes are observed regardless of $\Delta T$.}
\label{meanbondfluxes}
\end{figure}

Figure~\ref{meanbondfluxes} shows the $x-$, $y-$, and $z-$components of the 
mean bond flux $\langle\vec{\phi}\rangle$ as a function of the gradient span $\Delta T$ for $L=32$ and $T_1=0.5$.   
As expected from the symmetry along the transverse $x-$direction, there is no net bond flux in $x$: 
$\langle\phi_x\rangle=0$ regardless of $\Delta T$.   
For $\Delta T=0$, the system 
is also symmetric along $y$, so no net bond flux is observed. When a gradient is applied, however, 
this symmetry is broken. Since the growth probabilities depend on the Boltzmann factor 
$\exp(- \Delta E/T(y))$, where $T(y)$ is the layer's temperature, the thermal asymmetries introduced 
by the gradient favor spin deposition on the colder layers. This phenomenon is captured by the 
observed net bond flux $\langle\phi_y\rangle > 0$. Indeed, as shown in Figure~\ref{meanbondfluxes}, 
the thermal asymmetries cause $\langle\phi_y\rangle$ to grow steeply up to 
$\langle\phi_y\rangle_{max}\approx 0.75$ followed by a moderate decrease for larger gradients, 
which is due to the onset of bulk disorder within the hotter layers. 
Since the net transverse growth bond flux is directed from the ordered (cold) layers 
towards the disordered (hot) ones, 
this gradient-induced transverse ordering mechanism causes the system's critical temperature to increase 
from $T_c^{hom}=0.69(1)$ to $T_c=0.84(2)$. 
On the other hand, for $\Delta T=0$, $\langle\phi_z\rangle=1$ 
due to the longitudinal asymmetries in the substrate (i.e., the semi-infinite strip geometry constrains 
the system to grow along the $z>0$ direction). 
However, when the transverse gradient is applied, two effects contribute to decrease 
$\langle\phi_z\rangle$: $(i)$ the onset of the transverse bond flux, which creates transverse domains in the 
active perimeter and causes some of the added spins to grow backwards; $(ii)$ the bulk 
disorder induced in the hotter layers 
(which also causes $\langle\phi_y\rangle$ to decrease, as discussed above).

\begin{figure}[t!]
\center{\epsfysize=2.7in \epsffile{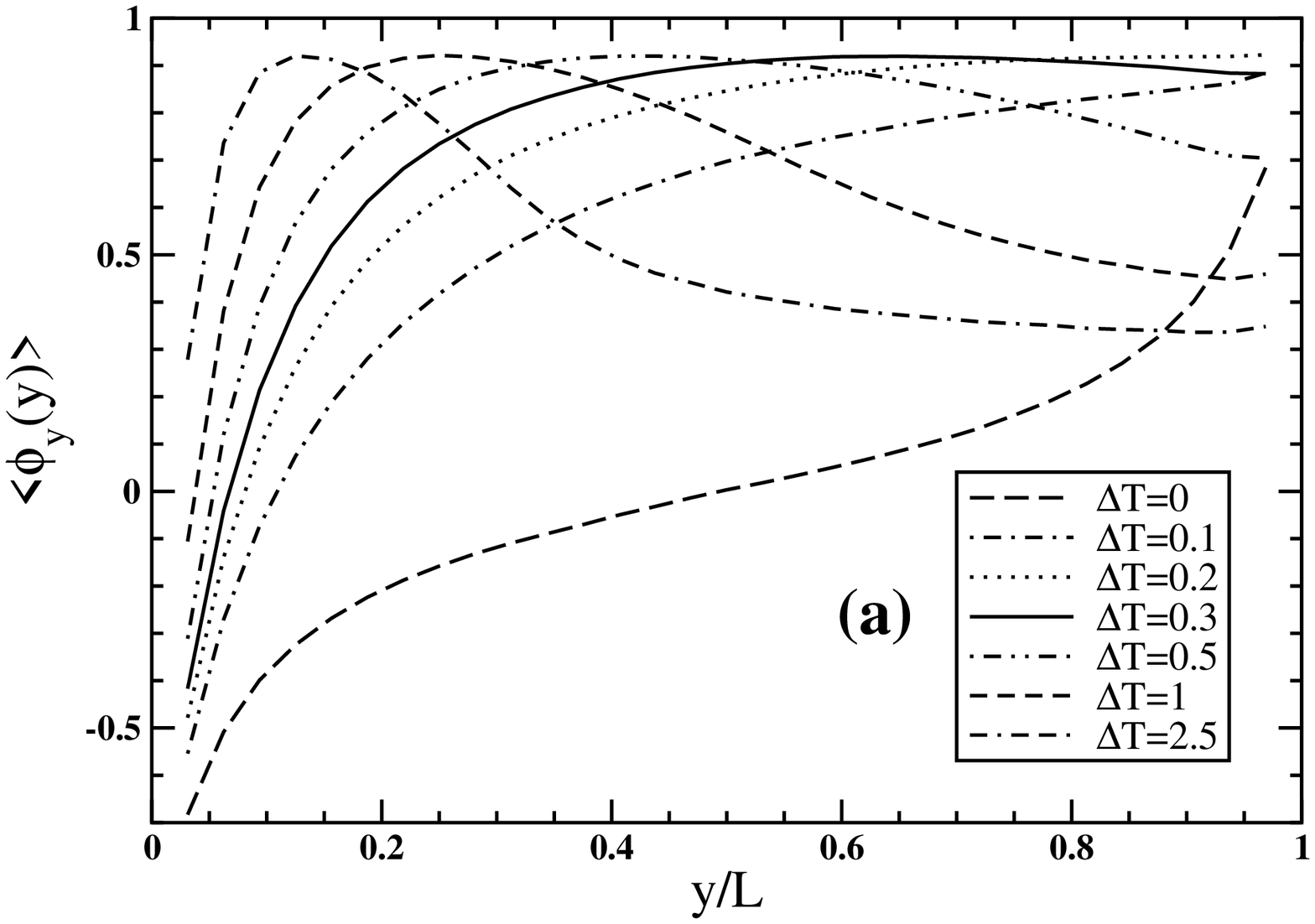}}
{\epsfysize=2.7in\epsffile{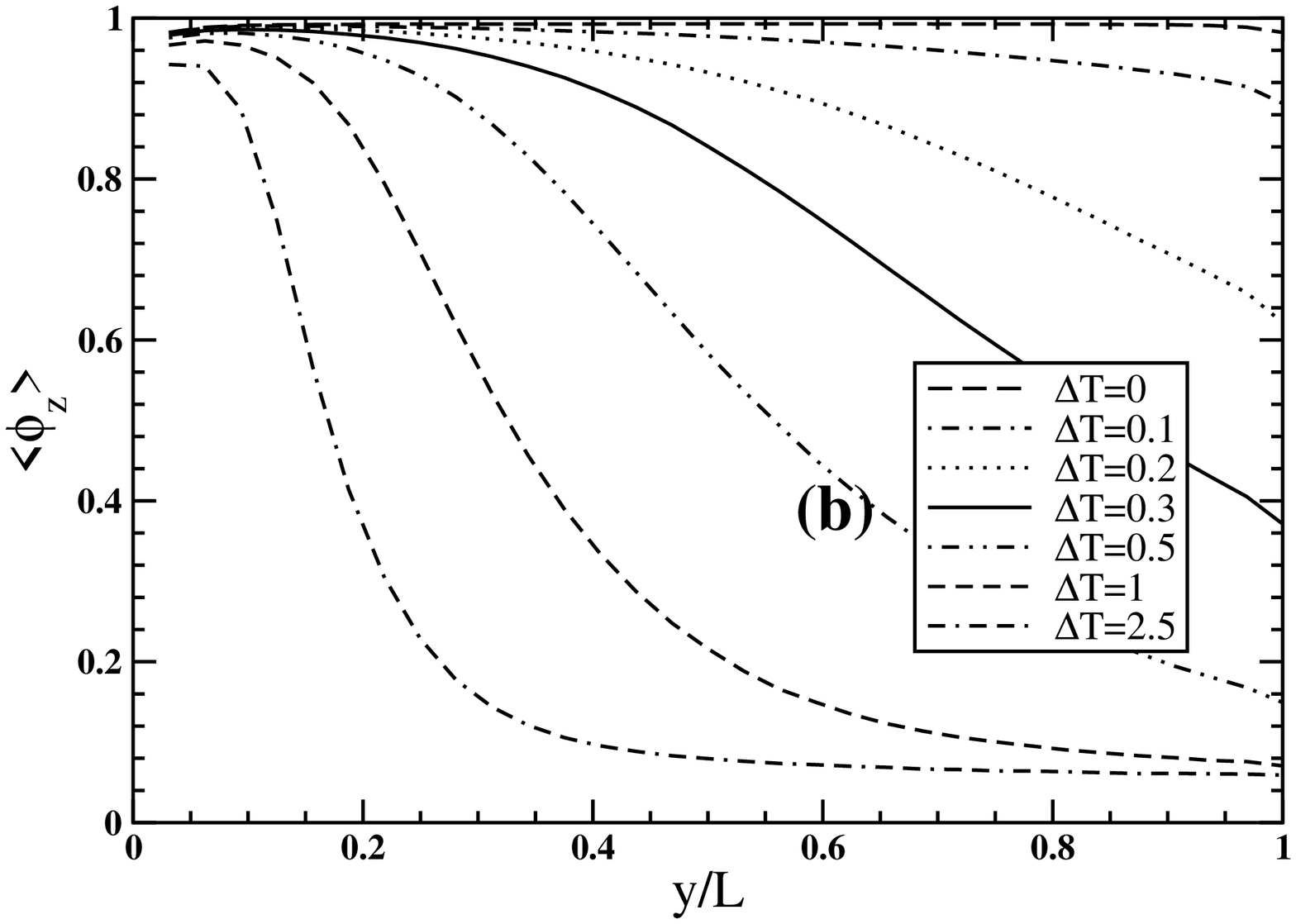}}
\caption{Bond flux components as a function of the transverse coordinate $y/L$ for $L=32$, $T_1=0.5$, 
and different gradient spans, as indicated: (a) transverse flux along $y$; (b) longitudinal flux along $z$.}
\label{bondfluxcomponents}
\end{figure}

The mean fluxes shown in Figure~\ref{meanbondfluxes} were averaged over sites at different temperatures. 
In order to gain further insight, let us now investigate the dependence of the bond fluxes 
on the layer temperature $T(y)=T_1+(y-1)\times(T_2-T_1)/(L-1)$ (where $1\leq y\leq L$). Since we are considering 
different gradient spans for a fixed system size, it is actually more convenient in this case to 
plot the bond fluxes as a function of the transverse coordinate $y$. 
We already observed that the transverse bond flux along $x$ is null regardless of temperature, so we will focus on 
the growth bond fluxes along the $y-$ and $z-$directions.
Figure~\ref{bondfluxcomponents}(a) shows the upwards 
bond flux $\langle\phi_y(y)\rangle$ vs $y$ for $L=32$, $T_1=0.5$, 
and different gradient spans $\Delta T$, as indicated. 
In the absence of a gradient, the flux is directed downwards at the bottom and upwards at the top, 
yielding zero net bond flux. Indeed, because of the open boundary conditions, empty perimeter 
sites at the confinement walls experience a missing-neighbor effect and the system grows 
preferentially along the center of the film as compared to the walls. 
When a gradient is applied, the flux grows steeply in the upwards direction, as expected. 
However, for larger gradients, the hotter thermal baths are capable of 
inducing disorder in the bulk and partially break the upwards bond flux on the upper layers. 
Indeed, this phenomenon causes the overall bond flux $\langle\phi_y\rangle$ (averaged over all 
layers) to decrease for large values of $\Delta T$, as discussed above. 
Similarly, Figure~\ref{bondfluxcomponents}(b) 
shows the forward bond flux $\langle\phi_z(y)\rangle$ vs $y$ for $L=32$, 
$T_1=0.5$, and different values of $\Delta T$, as indicated. For $\Delta T=0$, there is a slight 
missing-neighbor effect for the sites near the confinement walls. Since forward growth is mostly 
driven by the substrate asymmetry, this effect is much less noticeable that in the flux along $y$. 
When the gradient is applied, two effects contribute to reduce the longitudinal flux, as discussed 
above. One of them, which is dominant for small gradients, is due to the formation of transverse 
domains along $y$, causing some backwards deposition when the bulk is filled in. The other mechanism, 
which is dominant for larger gradients, is due to the onset of bulk disorder in the hotter layers.    

Previously, we pointed out the fact that the gradient-driven order-disorder phase transition occurs
at a temperature that is significantly higher than the corresponding critical temperature for 
the MEM growing in an homogeneous thermal bath (i.e. in the absence of a temperature gradient).
The results presented in this Subsection allow a qualitative 
explanation for the shift in the critical temperature. Indeed, the temperature gradient breaks the 
transverse symmetry along the $y-$direction, causing the onset of a 
net transverse growth bond flux. This flux is directed from the ordered layers (that grow at $T<T_c$) 
towards the disordered layers (that grow at $T>T_c$), thus expanding the low-temperature ordering 
effects across layers at higher temperatures. 
This gradient-induced transverse ordering mechanism increases the system's critical temperature. 

\subsection{Scaling behavior of the growth interface}
According to the analysis presented in Section 3.3, the system's critical temperature is increased 
due to gradient-induced transverse ordering mechanisms that originate in the cold layers. It could be 
argued, however, that, since the distance between the cold layer at $T_1$ and a layer at a 
fixed temperature $T>T_1$ becomes larger as $L$ is increased (while keeping fixed the 
gradient span $\Delta T$), then the transverse flux effects may become weaker and eventually negligible 
in the large-$L$ limit. In this Subsection we show that the growth interface is self-affine and its shape 
is stable and independent of size. Therefore, we confirm that the effects described in Section 3.3 are 
still relevant in the thermodynamic limit.  

\begin{figure}[t!]
\centerline{{\epsfysize=3.3in \epsfbox{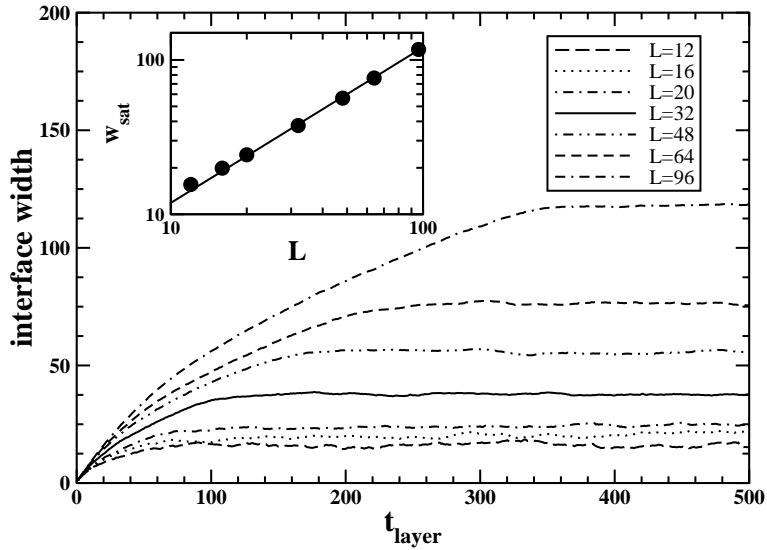}}}
\caption{Interface width as a function of time, where the time unit is the deposition time of $L^2$ spins. 
Inset: log-log plot of the saturation width, $w_{sat}$, as a function of system size. 
The statistical errors are smaller than the size of the symbols.
The fit to the data 
is shown by the solid line and agrees with a linear scaling $w_{sat}\propto L$ (more details in the text).
}
\label{transient}
\end{figure}

In order to track the evolution of the growth interface, we compute the position of the perimeter sites in 
the active region every time a monolayer of $L^2$ spins is deposited. The interface width at time $t$ is 
defined by
\begin{equation}
w(t)=\sqrt{{{1}\over{N_p}}\sum_{i=1}^{N_p}\left(z_i(t)-z_c(t)\right)^2}\ ,
\label{width}
\end{equation}
where the sum is taken over the $N_p$ perimeter sites in the active growth region, $z_i$ is the longitudinal 
coordinate of the $i-$th perimeter site, and $z_c=(1/N_p)\sum_i z_i$ is the center of the interface. 
Figure~\ref{transient} shows the interface width as a function of time for different lattice sizes in the 
range $12\leq L\leq 96$, where the time unit corresponds to the deposition of $L^2$ spins. After a short 
transient period, the interface reaches a stable saturation width, $w_{sat}$, analogously to other surface 
growth phenomena~\cite{vics92,bara95}. The Inset to Figure~\ref{transient} shows the dependence of $w_{sat}$ 
on the system size $L$, where the statistical errors are smaller than the size of the symbols. The fit to the 
data (solid line) 
shows that $w_{sat}\propto L^{\alpha}$, where the roughness exponent is $\alpha=1.01\pm 0.01$. That is, 
the saturation width scales linearly with the system size. 

\begin{figure}[t!]
\centerline{{\epsfysize=3.3in \epsfbox{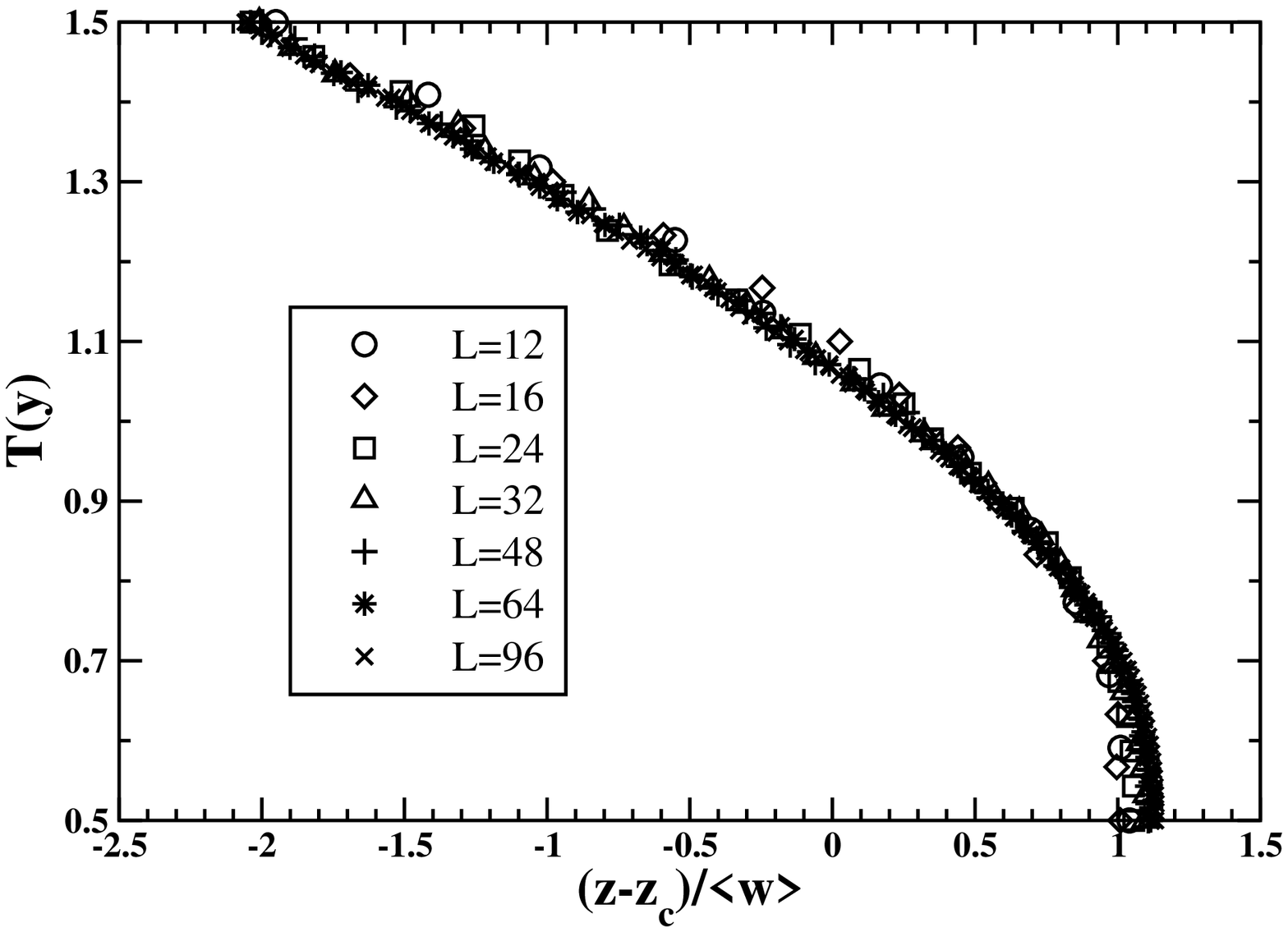}}}
\caption{Collapse of the scaled average growth profile in the stationary regime for different lattice sizes, as indicated.} 
\label{profile}
\end{figure}

By subtracting the interface center, $z_c$, we can compute the average interface profile in the stationary 
regime, as shown in Figure~\ref{profile}. We find that, when scaled by the mean interface width, interface 
profiles for different system sizes collapse into a universal shape for the growth interface. This nearly-linear 
universal shape is qualitatively consistent with the roughness exponent $\alpha = 1$,
as it has quantitatively been determined. In passing, 
notice also that this universal profile shows a good agreement with the instantaneous snapshot displayed in 
Figure~\ref{fig1}. Thus, we conclude that the active growth interface is self-affine and has universal 
features: the detailed analysis of bond flux asymmetries presented in Section 3.3 for a fixed lattice size 
($L=32$) remains valid for larger systems. In particular, our analysis focused on the influence exerted by the 
low-temperature layers into higher-temperature layers, which therefore is a gradient-induced growth 
mechanism still relevant in the thermodynamic limit. 

In addition to the roughness exponent $\alpha$, self-affine interfaces are also characterized by a fractal dimension $d_f$. 
In fact, within short length-scales such that $\Delta z\ll {\ell}$, where $\Delta z$ is the longitudinal interface distance 
along the growth direction between two points separated by a transverse 
distance ${\ell}$, the fractal dimension is $d_f = 2 - \alpha$~\cite{bara95}. In the long length-scale limit, moreover, the fractal dimension of the 
self-affine interface is $d_f = 1$, irrespective of its roughness~\cite{bara95}.   
Therefore, we conclude that the growth interface of the MEM in a thermal gradient is $d_f = 1$ at both short and long length-scales. 
Indeed, these conclusions are in full agreement with the nearly-linear shape of the growing interface that is consistent with a unitary fractal dimension at all length-scales, as seen in Figure~\ref{profile}.

Here, it is useful to compare the obtained results with those from related, well-studied growth models.   
For the standard MEM growing in a homogeneous temperature bath at high temperatures, the attachment of spins is a stochastic (random) process that becomes independent of the interaction energy and the temperature~\cite{ausl93}. 
Thus, in this limit, the growth interface of the standard MEM~\cite{ausl93,cand01,cand11} agrees with that of the classic Eden 
model~\cite{eden58}, which is well known to belong to the Kardar-Parisi-Zhang (KPZ) universality class~\cite{bara95}. The most accurate simulation 
results for the KPZ model in $(2 + 1)-$dimensions yield $\alpha= 0.393\pm 0.003$~\cite{mari00}, which agrees well with some of the formerly reported values 
for KPZ~\cite{amar90} and the Eden model~\cite{devi89}. Using this value for the roughness, the fractal dimension of the growth interface of the MEM 
in an homogeneous thermal bath at high temperatures crosses over from $d_f\simeq 1.6$ (at short length-scales) to $d_f=1$ (at long length-scales). 
Moreover, we can safely expect that the self-affine properties of the growth interface of the standard MEM (in an homogeneous bath) be independent of the temperature  within a wide range around and below the critical temperature, at least insofar the occurrence of a layering/roughening transition, in the sense of that observed in the $3-$dimensional Ising model~\cite{bind92,alex10}, can be neglected. Therefore, we conclude that the fractal and self-affine characteristics of the growth interface of the MEM in a constant temperature bath are quite different than those of the same model growing in a temperature gradient.
 
On the other hand, due to the fact that the MEM grown under a temperature-gradient constraint exhibits a second-order transition, one may also consider the self-affine properties of the interface between the ordered and the disordered phases. 
Although for systems under equilibrium conditions a useful (alternative)  approach is the evaluation of the damaged interface~\cite{bois91}, the damage spreading technique can not straightforwardly be applied for the evaluation of an interface in an irreversible growth model. 
Furthermore, the implementation and application of a cluster counting algorithm~\cite{sapo85, chap04, ross89, gouy05, ross86, losc09, guis11, ziff86, sabe08} 
to our model would require formidable computational task that is beyond the aim of this paper. Additional shortcomings for this kind of calculations are the definition of the suitable cluster (e.g. Swendsen-Wang vs physical clusters) and the occurrence of noticeable corrections to scaling that one needs to evaluate in order to obtain reliable exponents~\cite{chap04}, which is also a task that lies beyond our computational capabilities. 
However, from heuristic arguments based on the standard scaling relationship $\alpha_{ord-dis} = \nu /(1 + \nu) $~\cite{sapo85, chap04, bois91}, where $\alpha_{ord-dis}$ is the roughness exponent of the order-disorder interface, we can conjecture that $\alpha_{ord-dis} = 3/5$, which yields a self-affine 
order-disorder interface with short length-scale fractal dimension $d_f^{ord-dis} = 7/5$. For comparison, by applying the same scaling relationship to the 
standard MEM growing in an homogeneous thermal bath, we obtain $\alpha_{ord-dis}= 0.51\pm 0.09$ and $d_f^{ord-dis}= 1.49\pm 0.09$. 
Thus, although the growth interface is very significantly affected by the temperature gradient compared to the thermally homogeneous system, 
the geometry of the order-disorder interface is not so markedly affected by the gradient constraint and the results for both systems agree within 
error bars.  

\section{Conclusions}
In this work, we studied magnetic thin films growing under far-from-equilibrium conditions 
in $(2+1)$-dimensional strip geometries, where a temperature gradient is applied across 
one of the transverse directions. We modeled the thin film growth process by means of extensive Monte Carlo 
simulations performed on the magnetic Eden model (MEM), in which spins are deposited on a growing cluster 
with probabilities dependent on a ferromagnetic, Ising-like configuration energy. 
 
Firstly, we studied the thermal dependence of order parameter probability distributions, 
the order parameter mean absolute value (magnetization), the order parameter fluctuations 
(susceptibility) and its higher moments (Binder cumulant) on finite-size magnetic films, which 
showed the existence of gradient-driven pseudo-phase transitions.  
Secondly, we applied Binder's cumulant method and finite-size scaling analysis in order 
to characterize quantitatively the critical behavior of MEM films growing in a temperature gradient. 
The system's critical temperature 
is $T_c=0.84(2)$, significantly higher than the MEM's critical temperature when growing in an homogeneous 
thermal bath, namely $T_c^{hom}=0.69(1)$~\cite{cand01}. The critical exponents are $\nu=1.53(6)$,  
$\gamma=2.54(11)$, and $\beta=0.26(8)$, 
which also differ from the MEM's exponents in the absence of a temperature gradient~\cite{cand01}. 
By changing the gradient span, we observed finite-size effects that vanish in the thermodynamic limit.
Hence, the critical temperature and exponents are universal for MEM films growing in a temperature gradient. 
We also investigated the system's growth dynamics by means of a bond model. We found that 
the interplay of geometry and thermal bath asymmetries leads to growth bond flux 
asymmetries and the onset of transverse ordering effects that explain qualitatively 
the shift observed in the critical temperature. Finally, we analyzed the self-affine growth interface
and obtained a collapse of the scaled average growth profile in the stationary regime for different lattice sizes, which shows that growth bond flux asymmetries play a relevant role in the model's growth dynamics 
even in the thermodynamic limit. 

In the context of a great experimental and theoretical interest in magnetic systems growing in 
temperature gradients, 
as well as a wide variety of technological applications that benefit from these efforts, 
we hope that this work will contribute to the progress of this research field and stimulate further work.  

\section*{Acknowledgments}
This work was financially supported by CONICET, UNLP, and ANPCyT (Argentina). The authors thank Nara Guisoni 
and Ernesto Loscar for useful discussions.

\end{document}